# Opportunities and Enabling Technologies for 5G Spectrum Sharing

Maziar Nekovee,
Deprtment of Engineering and Design,
School of Engineering and Informatics , University of Sussex
Brigton BN1 9RH, UK

m.nekovee@sussex.ac.uk

**Abstract.** In this paper an overview is given of the current status of 5G industry standards, spectrum allocation and use cases, followed by initial investigations of new opportunities for spectrum sharing in 5G and the underlying technologies to enable efficient sharing, considering both licensed and unlicensed scenarios and spectrum both below 6 GHz and in the millimeter-wave frequency range.

## 1 What is 5G?

5G is the next generation of mobile communications technology and is being designed to provide (in comparison with 4G) greater capacity, faster data speeds, and offer very low latency and very high reliability, enabling innovative new services across different industry sectors. The first wave of 5G commercial products is expected to be available in 2020 although some "pre-5G" deployments are already expected in 2018. 5G technology standards are currently under development, and will include both an evolution of existing (4G) and new radio technologies (5G NR).

### 1.1 5G use cases

According to the International Telecommunications Union ITU [ITU] who has defined vision and requirements of IMT2020 [1], potential 5G services and applications can be grouped into three different classes:

• Enhanced Mobile Broadband. Together with an evolution of the services already provided by 4G, 5G is expected to provide much faster and more reliable mobile broadband, offering a richer experience to consumers for application such a virtual reality (VR) and augmented reality (AR) as well as cloud-based services. The specific requirements are a minimum of 100 Mbps user-experience data rates and 20 Gbps peak data rate.

• Massive Machine Type Communications. The Internet-of-Things (IoT) – where sensors. actuators, consumer electronics appliances, street lighting etc wirelessly connect to the internet and each other. This is already happening on existing 4G networks and the technology is being used in everything from smart homes to wearables. 5G should help the evolution of IoT services and applications and improve interaction between different platforms as well as enabling the vision of 50 billion devices becoming connected by 2030. Possible future applications could include real-time health monitoring of patients; optimisation of street lighting to suit the weather or traffic; environmental monitoring and smart agriculture. Data security and privacy issues will need to be considered given huge amounts of data could be transferred over a public network. We note that many IoT services are already being offered or will be offered in the next few years over existing and evolved 4G networks , e.g. using Narrow-Band IoT (NB-IOT), LTE-M or NB-LTE-M technologies. 5G, in this area, is likely to kick-in by about 20205 where we expect to see the through explosion of new IoT services for which evolution of LTE is unable to address the required scalability requirements.

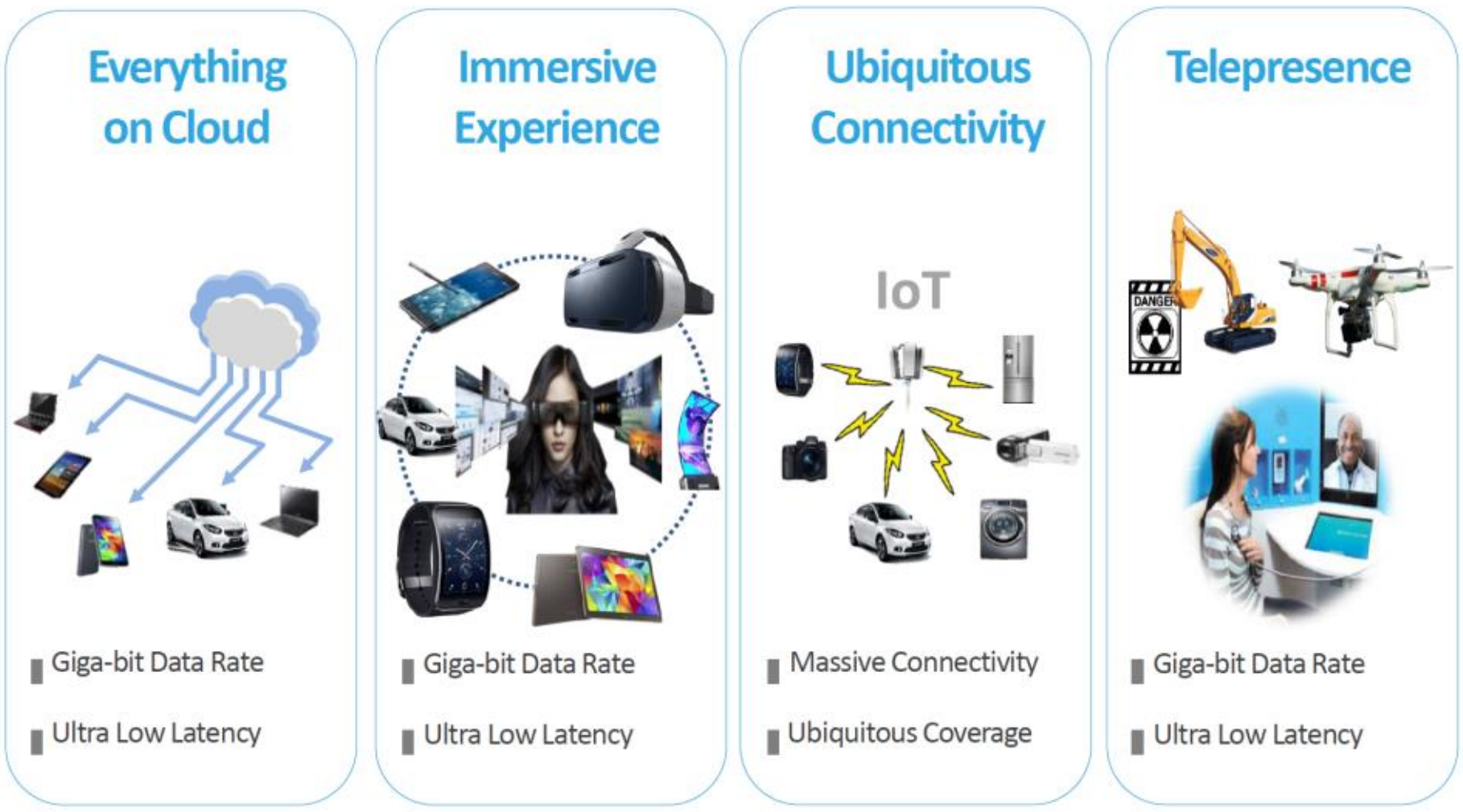

*Figure 1 Exemplar Use cases of 5G and their requirements.*

- Ultra-reliable and low latency communications. This class is likely to rely on the new radio developments, and includes services requiring a very high reliability and/or a very low latency. Possible applications include Connected and autonomous cars and aerial vehicles, remote control of robots in extreme conditions and hazardous situations and for industry automation (Industry 4.0), remote surgery and the so-called tactile Internet as well as some of the applications in the context of smart grids.

These different services have different requirements in terms of speed, coverage, latency and reliability, which will demand different network solutions (the evolution of existing network and potentially new networks) and different deployment models (including many small cells), an appropriate network infrastructure (which will include both fibre and wireless connectivity to the core network) and access to different spectrum bands. Therefore, the concept of network slicing is being put forward, where different slices of the overall 5G network infrastructure (including spectrum) may be allocated to different types of services to end-users.

We note that in addition to the above three categories of 5G use cases, perhaps surprising a new use case for 5G, the so-called Fixed Wireless Access (FWA) or fibre-like wireless has recently emerged [2,3]. FWA refers to the provision of high data rate (> 100 Mbps) broadband wireless access to residential customers and enterprise premises using pre-5G/5G access technologies, including Full-Dimensional MIMO (FD-MIMO), Massive MIMO and millimetre-wave radio access technologies. The FWA concept has been around for quite a long time (being known also as wireless local loop) but only with 5G the techno-economic case for this use case has become a compelling alternative to wired solutions, such as next generation cable, cooper-based G-fast and Fibre-to-Premises (FTTP).

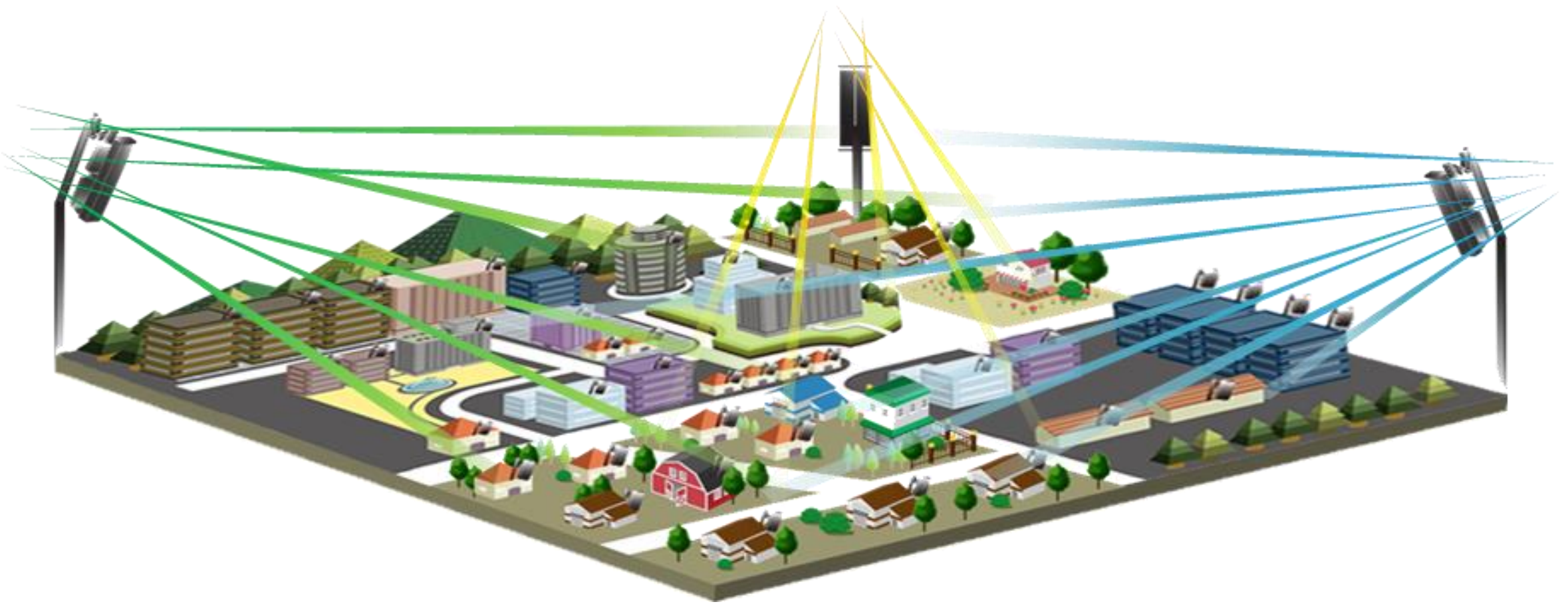

***Figure 2 5G Fixed Wireless Access architecture.***

### 1.2 Radio access technologies for 5G

Different from previous generations, where a new radio access technology replaced the old one, 5G will integrate different radio technologies. Some of these will be the evolution of already existing radio access technologies, some will be new. Different service classes could rely on different radio interfaces. Evolutions of the latest version of the 4G radio interface (LTE-Advanced Pro) are likely to be used to provide a coverage layer via macro cells. A new cellular radio interface (being developed in 3GPP under the name 'New Radio' or 'NR') operating in frequencies up to 50 GHZ will be used to provide very high data rates, ultra low latencies and to serve a very large number of devices via a large number of small cells. Low-cost, low-battery consumption IoT services are likely to be delivered initially using evolved 4G technologies, as described in the Introduction, with a gradual transition/phase-out to 5G by 2025.Wi-Fi evolutions will also play an important role for consumers, in particular to provide 5G services within homes or offices. In addition it is also expected that Satellite technologies also play a role in 5G, in particular for wide area coverage in IoT application space (e.g. tracking of goods and vehicles ) , and also as mechanism to offload broadcast and multicast linear TV traffic from 5G cellular networks.

### 1.3 5G standards timelines

Figure 3 shows the latest (as of 25/06/2017) timeline of 3GPP (Third Generation Partnership Project), which is responsible for developing a global industry standard for 5G mobile communication technologies. As can be seen from this figure, 5G phase 1 standards, which are mainly focusing on enhance Mobile Broadband (eMBB) with some element of ultra-low-latency included, are expected to be ready mid 2018, with an initial "non-standalone" version of the standard to be released already by the end of 2017. The second phase of 5G technology, which encompasses massive machine-type communications and ULL, is expected to be ready by the end of 2019, in time for the standard to be proposed to ITU as a candidate technology which fulfils the IMT 2020 requirements.

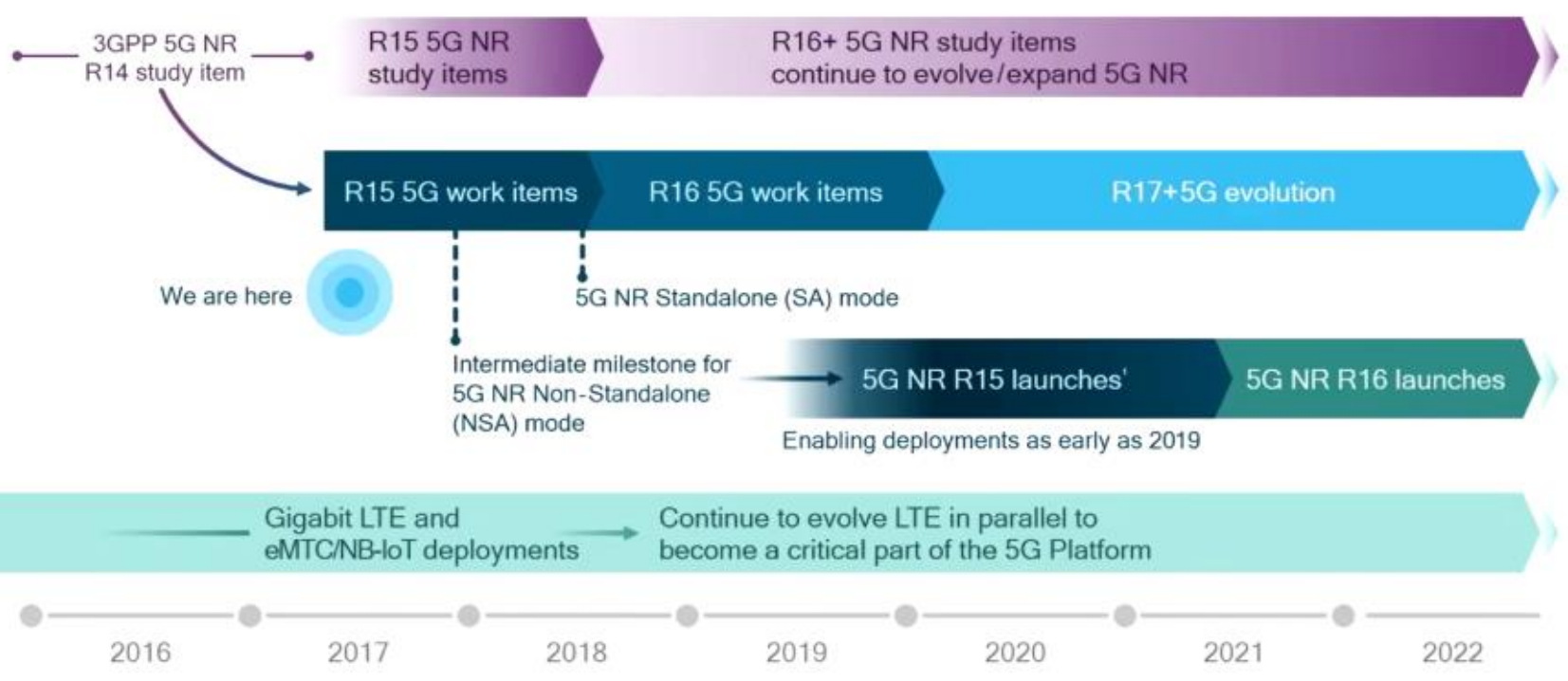

Figure 2 5G standardization timeline according to 3GPP.

## 2 Spectrum for 5G

Spectrum is a critical component of wireless networks. It makes up the 'airwaves' that underpin the communication services we use every day; such as mobile, Wi-Fi and TV. The diverse set of 5G services and applications, described above, will require a diverse set of spectrum bands, with different characteristics, addressing different requirements, and combining both low and high frequencies:

- Spectrum at lower frequencies, and in particular below 1 GHz, to enable 5G coverage to wide areas;
- Spectrum at higher frequencies with relatively large bandwidths below 6 GHz , to provide the necessary capacity to support a very high number of connected devices and to enable higher speeds to concurrently connected devices; and

- Spectrum at very high frequencies above 24 GHz (e.g. millimetre wave) with very large bandwidths, providing ultra-high capacity and very low latency. Cells at these frequencies will have smaller coverage (between 50-200 m) and it is likely that build-out of 5G networks in millimetre wave bands will initially be focused on areas of high traffic demand, or to specific locations or premises requiring services with very high capacity and/or peak data rates (Gbps).

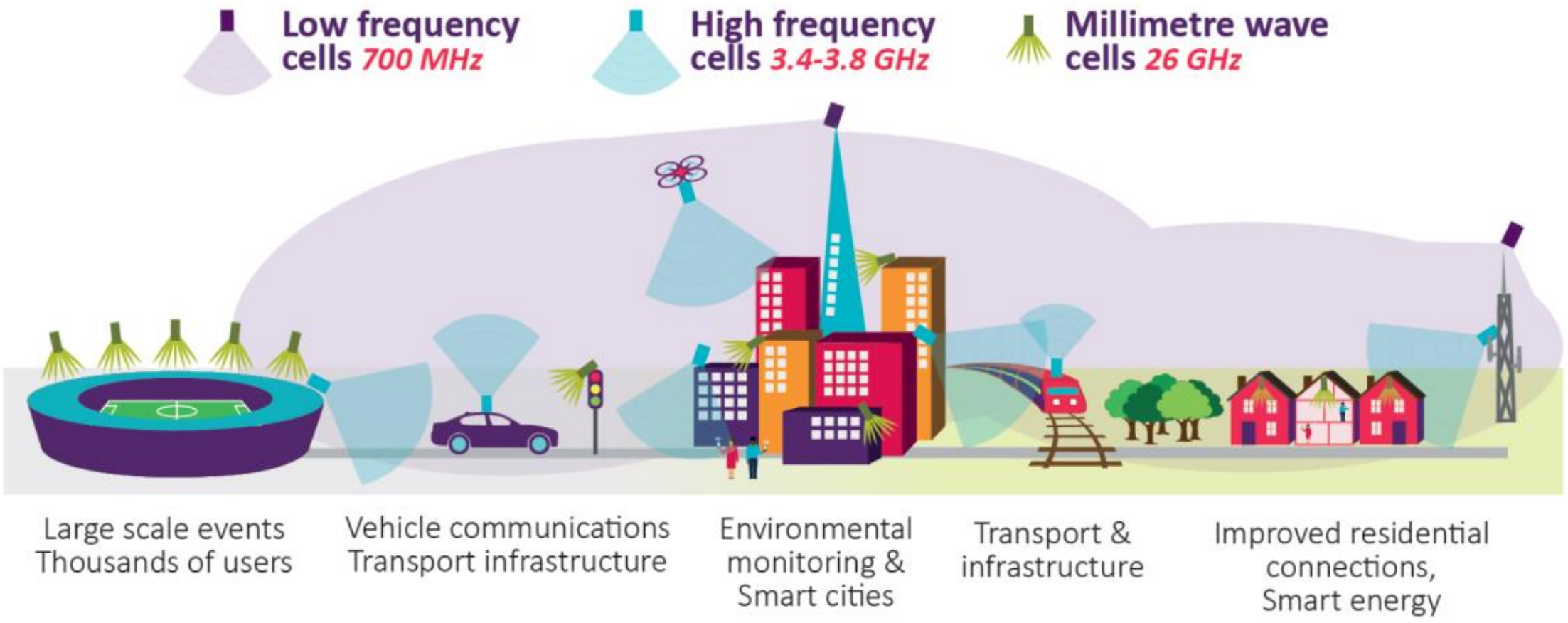

*Figure 3 Radio spectrum for 5G and its uses [Ofcom].*

### 2.1 5G spectrum allocation

The 2015 World Radio Congress (WRC-15) agreed on a WRC-19 Agenda Item (1.13) to consider the identification of frequency bands for the future development of International Mobile Telecommunications (IMT), including possible additional allocations to the mobile service on a primary basis, in accordance with Resolution 238 (WRC-15). This involves conducting and completing the appropriate sharing and compatibility studies for a number of bands between 24-86 GHz in time for WRC-19. The compatibility and sharing studies for these bands are being carried out in ITU-R Task Group 5/1 until 2018. This follows work in ITU-R on spectrum needs, deployment scenarios, sharing parameters and propagation models which were completed in March 2017.

Candidate bands identified for further study in WRC-15 are shown in Figure 3. It can be seen that there are regional differences and in particular in the 20-30 GHz range it can be expected that the 27.5-29.5 band will be only available in Americas while other regions, including Europe, are likely to converge around the 24.25-27.5 range.

**Frequency Ranges Below/Above 6 GHz of by Region (WRC-15)**

Below 6GHz | Above 6GHz/mm-Wave

| | < 6GHz (MHz) | 6-20 | *20-30* | 30-40 | 40-50 | 50-60 | 60-70 | 70-80 | 80-100 |
|---|---|---|---|---|---|---|---|---|---|
| ***APAC*** *(APT)* | 1427 – 1452<br>1492 – 1518 | | 25.25 – 25.5 | 31.8 – 33.4 | 39 – 47<br>47.2 – 50.2 | 50.4 – 52.6 | 66 – 76 | | 81 – 86 |
| ***Europe*** *(CEPT)* | 1427 – 1518<br>3400 – 3800 | | 24.5 – 27.5 | 31.8 – 33.4 | 40.5 – 43.5<br>45.5 – 48.9 | | 66 – 71<br>71 – 76 | | 81 – 86 |
| ***Americas*** *(CITEL)* | 1427 – 1515<br>3488 – 3600 | 10 – 10.45 | 23.15 – 23.6<br>24.25 – 27.5<br>**27.5 – 29.5** | 31.8 – 33<br>37 – 40.5 | 45.5 – 47<br>47.2 – 50.2 | 50.4 – 52.4 | 59.3 – 76 | | |
| ***Russia*** *(RCC)* | 5925 – 6425 | | 25.5 – 27.5 | 31.8 - 33.4<br>39.5 - 40.5 | 40.5 – 41.5<br>45.5 – 47.5<br>48.5 – 50.2 | 50.4 – 52.4 | 66 – 71<br>71 - 76 | | 81 - 86 |
| ***Mid. East*** *(ASMG)* | 1452 – 1518<br>3400 - 3600 | | | 31 - 100 | | | | | |

※ APT : Asia-Pacific Telecommunity (APT)
CITEL : Inter-American Telecommunication Commission
ASMG : Arab Spectrum Management Group
CEPT : European Conference of Postal and Telecommunications Administrations
RCC : Regional Commonwealth in the Field of Communications (Russia etc.)

Figure 4 Candidate frequency bands for 5G as identified in WRC15.

In parallel on the European level Radio Spectrum Policy Group (RSPG) has developed in 2016 a strategic roadmap for 5G in Europe. In particular, the roadmap identified the following building blocks for 5G:

Low bandwidth spectrum at **700 MHz**; Medium bandwidth spectrum at **3.4 – 3.8 GHz** as a "primary" band, which will provide capacity for new 5G services; and
• High bandwidth spectrum at **24.25 – 27.5 GHz** as the "pioneer" millimetre wave band to give ultra-high capacity for innovative new services, enabling new business models and sectors of the economy to benefit from 5G. In addition, A European Commission Mandate to CEPT was approved by Member States with regards to the development of harmonised technical conditions in two "pioneer" bands: 3.4 to 3.8 GHz and the 26 GHz band.

# 3 5G Spectrum Sharing

## 3.1 Sharing below 6 GHz spectrum

While above 6 GHz large chunks of spectrum are expected to become available for 5G systems, the amount of spectrum the sub-GHz and below 6 GHz range is far more limited. The sub-6GHz band is expected to support important applications of 5G, such a machine-type communications due to excellent propagation and indoor penetration characteristics while the first wave of 5G mobile communication systems are expected to be deployed in 3.6 GHz frequency range, where in conjunction with the use of Massive MIMO and Full-dimensional MIMO (FD-MIMO) technologies 100Mbps+ data-rates could be supported while also keeping cell-sizes sufficiently large for viable deployment. It is, therefore of great importance to explore options for sharing of these very precious portions of 5G spectrum.

Due to quality of service requirement of 5G use cases that are expected to be supported, a very important option for sharing of these bands is the evolution of Licensed Shared Access (LSA) [5]. In this approach licensed users, called LSA licensees, can access underutilized licensed spectrum on an exclusive basis, thus enjoying predictable QoS, when it is not being used by the incumbent, hence protecting it from harmful interference.

In 2017, the Federal Communications Commission (FCC) opened up 150 MHz of spectrum in the U.S. around 3.5 GHz that it named Citizens Broadband Radio Service (CBRS) [cbbs]. Similar to the LSA approach, CBRS enables others to use the spectrum while it is still being used by existing incumbents, such as the military or satellite communication, see Figure 7. However, in addition to sharing with incumbents — CBRS adds a 'third-tier' of general usage. In this third-tier, anyone can use the spectrum when it is not used by the higher tiers (the incumbents or users that paid for a license), see Figure 2. Of course, if there are multiple third-tier users in the same area then they will share the available spectrum with each other in a fair manner. The complexity of managing three tiers will require some additional control. To this effect, the FCC has defined a Spectrum Access System (SAS) — a type of database, in effect — and the Wireless Innovation Forum is helping to specify the details to ensure that it all works in accordance with the FCC rules.

CBBS can be used by existing mobile operators to offer Gigabit LTE in more places by making more spectrum available. One can also use this spectrum for small-cell deployments to extend coverage and add capacity indoors. Another foressen use is the so-called neutral host, which is a LTE deployment that can be used by subscribers irrespective of their service provider.

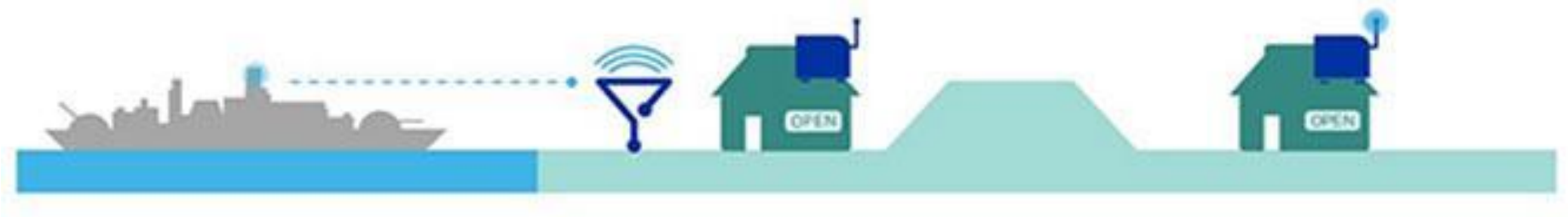

**Figure 5 CBBS spectrum sharing scenario in 3.5 GHZ between an incumbent system (in this case Navy radar) and secondary users.**

### 3.2 Sharing mm-wave spectrum

Millimetre-wave (mm-wave) communications have emerged as a key disruptive technology for both cellular networks (5G and beyond) [6] and wireless Local Area Networks (802.11ad and beyond). While spectrum availability is limited in traditional bands below 6 GHz, mm-wave frequencies offer order of magnitude greater bandwidths. In addition, mm-wave communication is typically characterized by transmissions with very narrow beams, enabling further gains from directional isolation between mobiles. This combination of massive bandwidth and spatial degrees of freedom may make it possible for mm-wave to meet some of the boldest 5G requirements, including higher peak per-user data rate, high traffic density and very low latency. The use of mm-wave bands for 5G present a number unique features not present at lower frequencies:

- **Beamforming as a mandatory requirement**: A common characteristic of all systems operating in mm-wave frequencies is that beam-forming is mandatory to compensate for the significantly higher pathloss in these frequencies. For example the IEEE 802.11ad standard supports up to four transmitter antennas, four receiver antennas, and 128 sectors. Beam forming is mandatory in802.11ad, and both transmitter-side and receiver-side beamforming are supported. Furthermore specification of beamforming for 5G are expected to be finalized by 3GPP, as part of 5G New Radio (NR) work item. Consequently, beams provide a common new dimension for sharing of spectrum among multiple access technologies.
- **Potential for "infinite" spatial reuse:** Wireless communications systems already rely on spatial sharing of spectrum in two dimensions, and the entire concept of cellular communications relies on spatial re-use of radio spectrum. In mm-wave systems with the use and transmit-side and receive-side beamforming spatial spectrum re-use can be pushed even further down close to one-dimension, which the footprint of interference from each transmission link becoming very close to line, rather than an area, in two dimension. In the idealized case of ultra-narrow beams this would allow infinite spatial reuse of spectrum.

**Sharing with satellite services**

FSS (Fixed Satellite Service) is the official classification for geostationary communications satellites that provide, for instance, broadcast feeds to television stations, radio stations and broadcast networks. The FSS uplink (from FSS to satellite) is allocated in the band from 27.5 to 30 GHz, which is adjacent to the 24.25-27.5 GHz band identified for 5G. Therefore, there could be potential issues with sharing between 5G and FSS due to adjacent channel interference. Several cognitive techniques can be applied to mitigate improve the 5G-FSS coexistence.

The coexistence between FSSs and mobile cellular BSs in the mmWave bands have been the subject of only few, and mainly theoretical, studies. Important new parameters that need to be considered are how the interference levels could be reduced by exploiting multiple antenna configurations by 5G mm-wave systems as well as investigating the aggregate interference resulting from massive deployment of 5G systems on uplink FSS. The studies in [7, 8], performed in the wosre-case scenario of co-channel sharing have indicated that due to the use of a beamforming technology, combined with the relatively short-range of communications in mm-wave frequencies, spatial sharing is much more feasible than in the case of IMT-advanced systems.  In particular, even in this worse-case scenario the required protection distance around FSS are much smaller (~1km as opposed to hundreds of km) than those recommended previously. Furthermore, by using coordination among multiple 5G BS further gains in spectrum sharing can be achieved. These studies also indicate that presence of highly directional FSS transmission can  cause outage in the coverage of 5G mm-wave network. However, due to the highly directional FSS transmission, the outage region is well-confined and its impact could be mitigated using a combination of null-forming at 5G UE's and cooperation by multiple BS to boast signal strengths at the victim UE.

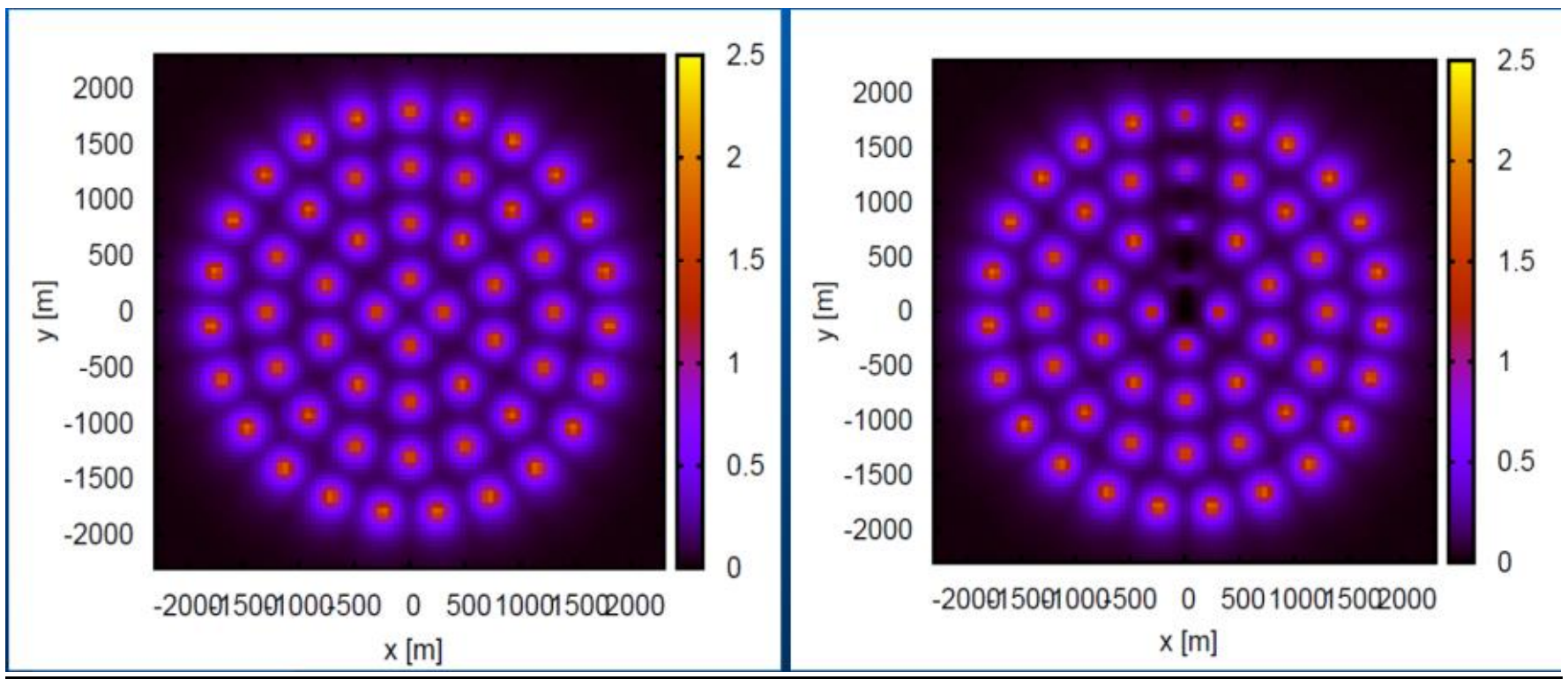

***Figure 6 Impact of FSS uplink transmission on the coverage of a mm-wave 5G network in the worse-case co-channel sharing of 28 GHz spectrum. Coverage maps are shown in the absence (left panel) and presence (right panel) of a FSS's highly directional  transmitter positioned at the centre of the area.***

**Sharing between access and backhaul (fixed) links**

The 26 GHz frequency range is used in many countries for 4G backhaul links, also known as fixed links (FL). Therefore it is of  prominent importance to investigate the compatibility of using the the 24.25-27.5 GHz for 5G access.  A recent study of coexistence between FL and 5G access links has been performed by Ofcom [ofcom]. The study considered the deployment of 5G mm-wave base stations (BS) in London overlaid on the existing FS deployment.  It was assumed that, to avoid harmful interference to FS links, 5G BS can only operate if they are outside the denial area of FS links. Denial areas where assumed to be circular and were derived from interference analysis of a single 5G BS on a single FS link.

The analysis shows that, in the non-LoS case, 5G BS would not be able to be rolled-out within 0.5 km of a typical FL without a 20% probability of causing unacceptable levels of interference in all directions. This probability significantly increases in the pointing direction of the FL, while the distance at which a 20% probability of causing unacceptable levels of interference increases to 1.2 km. Therefore, to ensure that unacceptable interference was not caused to this incumbent FL, this area would have to be denied to IMT-2020 BSs. In fact, dependent on the requirements of the incumbent FL, and what probability of interference they were willing to accept, this denial area may need to be significantly larger.

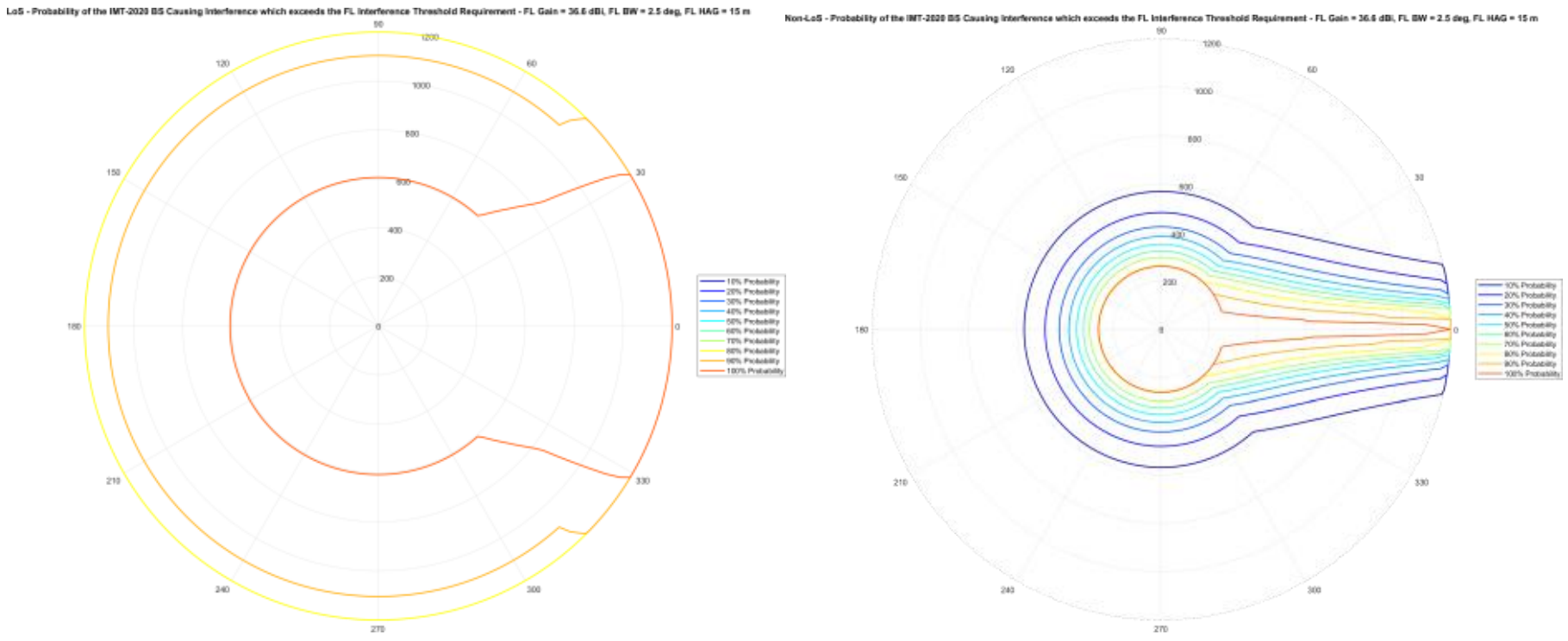

**Figure 5 Contour plots showing the probability of a 5G base station causing harmful interference to a fixed link operating in 26 GHz assuming line-of-sight (left panel) and non-line-of-sight (right panel) results.**

Figure 6 shows what impact this denial area would have on a mobile operator wanting to deploy 5G BSs in an urban conurbation, provided they were willing to ensure that their 5G BS did not have LoS to any FL. In this figure a denial area of 1.2 km is placed over all FLs in London which overlap with the block of 400 MHz of spectrum from 24.5 GHz to 24.9 GHz (shown in red).

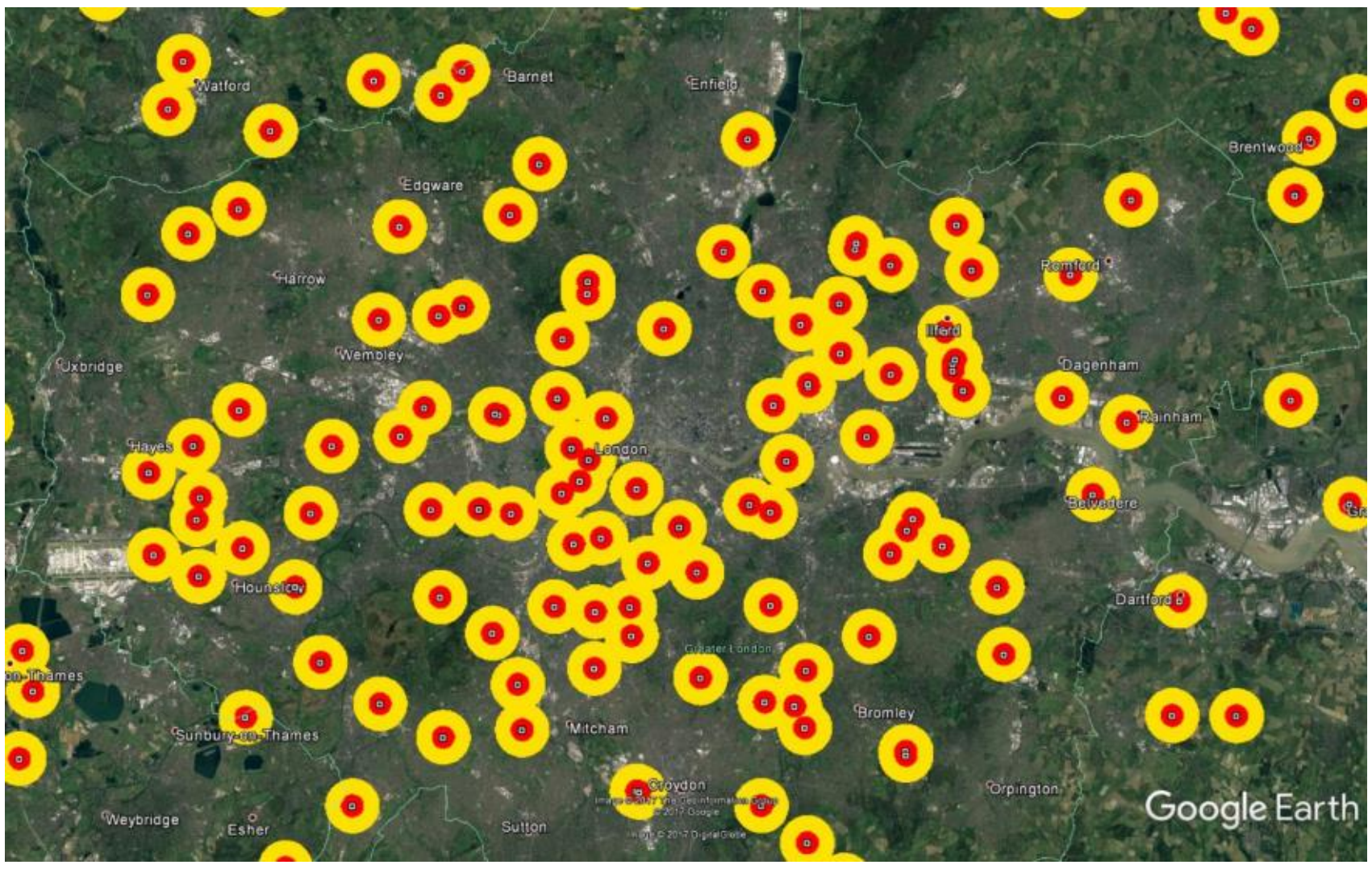

**Figure 7 Denial area within 1.2km of all backhaul links in London operating in the 24.9 GHz to**

## Sharing unlicensed mm-wave spectrum

A recent trend in cellular communication is to utilize both the licensed and unlicensed spectrum simultaneously for extending available system bandwidth. In this context, LTE in unlicensed spectrum, referred to as LTE-U, is proposed to enable mobile operators to offload data traffic onto unlicensed frequencies more efficiently and effectively, and provides high performance and seamless user experience. Integration of unlicensed bands is also considered as one of the key enablers for 5G cellular systems. However, unlike the typical operation in licensed bands, where operating base stations (BS) have exclusive access to spectrum and therefore are able to coordinate by exchanging of signalling to mitigate mutual interference, such a multi-standard and multi-operator spectrum sharing scenario (as shown in Figure 4) imposes significant challenges on coexistence in terms of interference mitigation. Licensed Assisted Access (LAA) with listen-before talk (LBT) protocol has been proposed for the current coexistence mechanism of LTE-U. In case of mm-wave unlicensed sharing a major issues is that the use of highly directional antennas as one of the key enablers for 5G networks becomes problematic for the current coexistence mechanisms where omni-directional antennas were mostly assumed. For example, as shown in Fig. 7 transmission by a different nearby 5G BS or WiGig access Point (AP) may not be detected due to the narrow beam that has been used, resulting in “beam-collision” which can cause even more excessive interference than in conventional systems.

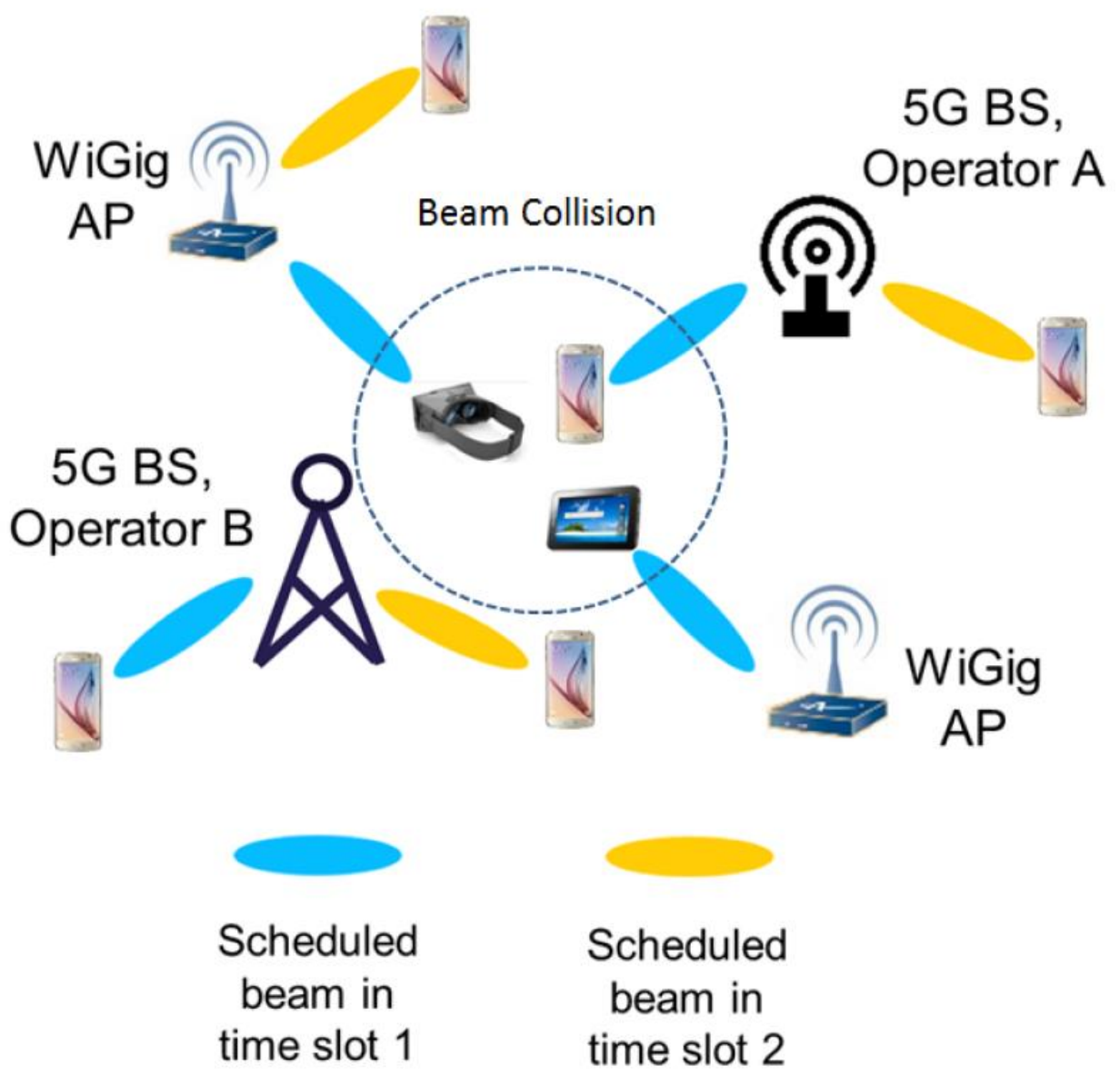

*Figure 8 Multi-standards and multi-operator sharing of unlicensed mm-wave bands in future 5G systems, showing also a beam collision interference scenario [10].*

We note that such beam-collision interference scenarios can also occur in exclusively used mm-wave spectrum as well. However in such scenarios, centralized resource allocation algorithms from 4G can be extended to include beam-scheduling among multiple base stations to avoid such excessive interference scenarios. In the case of unlicensed sharing of mm-wave spectrum, centralized coordination is not possible, and novel mechanisms need to be developed. Work in this direction has only recently being started as part of a new study item in 3GPP 5G-NR which is expected to be completed in 2018 [10,11]. Various mechanism for sharing are being proposed, including distributed and self-organized mechanism for beam-coordination [nekovee1, nekovee2], and approaches based on spectrum pooling [boccardi].

## 4 Conclusion

With industry standards for 5G cellular systems rapidly progressing and firming up, issues and challenges related to future sharing and coexistence of spectrum are starting to take a center stage. Furthermore, there is a strong desire from governments and regulators for efficient allocation and use of 5G spectrum. Therefore, given also the maturity of technologies such as LSA, LLA, cognitive radio and mm-wave communications, we can expect that spectrum sharing will be a very prominent area for innovation, standardization and spectrum regulation in the next few years and beyond.

## References

1. Recommendation ITU-R M.2083-0, IMT Vision – Framework and Overall Objectives of the Future Development of IMT for 2020 and Beyond, September 2015.
2. Ovum (Sponsored by Samsung), 5G Fixed-Wireless Access, Providing Fiber Speeds Over the Air While also Helping Pave the Way for 5G Full Mobility, 2016
3. Laraqui K., Tombaz S., Furuskär A. , Skubic B., Nazari A., Trojer E., Fixed Wireless Access on Massive Scale for 5G , Ericsson Technology Review December 2016.
4. ITU-R Working Party 5D, Minimum requirements related to technical performance for IMT-2020 radio interface(s), February 2017.
5. Matinmikko, M., et al. "Spectrum sharing using licensed shared access: the concept and its workflow for LTE-advanced networks. IEEE Wireless Communications Magazine (2014): 72-79.
6. Roh W., Seol J-Y, Park J., Lee B., Lee J., Kim Y., Cho J., Cheun K, Aryanfar F., Millimeter-wave beamforming as an enabling technology for 5G cellular communications: theoretical feasibility and prototype results. IEEE Communications Magazine 52(2): 106-113 (2014)
8. Guidolin F., Nekovee M., Investigating Spectrum Sharing between 5G Millimeter Wave Networks and Fixed Satellite Systems. IEEE GLOBECOM Workshops 2015: 1-7
9. Guidolin F., Nekovee, Badia L., Zorzi M., A study on the coexistence of fixed satellite service and cellular networks in a mmWave scenario. IEEE ICC 2015: 2444-2449
10. Nekovee M., Qi Y., Wang Y., Self-organized Beam Scheduling as Enabler for Coexistence in 5G Unlicensed Bands, IEIC Trans. 2017.
11. 3G PPP TSG RAN Meeting #75, RP-170828, Study on NR-based Access to Unlicensed Spectrum, March 2017.